\documentclass[12pt,a4paper]{article}
\usepackage{a4wide}
\usepackage{amsmath}
\usepackage{amssymb}
\usepackage{amsfonts}
\usepackage{ifpdf}
\usepackage{graphicx,epsfig}
\usepackage{subfigure}
\usepackage{exscale}
\usepackage{float}
\usepackage{bbm}
\usepackage[numbers,sort&compress]{natbib}

\usepackage{multicol,multirow}

\newcommand{\R}{{\mathbb{R}}}

\newcommand{\evec}{\vec{e}} 

\def\lsi{\raise0.3ex\hbox{$<$\kern-0.75em\raise-1.1ex\hbox{$\sim$}}}
\def\gsi{\raise0.3ex\hbox{$>$\kern-0.75em\raise-1.1ex\hbox{$\sim$}}}
\newcommand{\lsim}{\mathop{\lsi}}
\newcommand{\gsim}{\mathop{\gsi}}

\makeatletter
\@addtoreset{equation}{section}
\makeatother

\title{\vspace{3mm} Topological Lattice Actions for the 2d XY Model}

\author{W.\ Bietenholz$^{\rm a}$, M.\ B\"ogli$^{\rm b}$, 
F.\ Niedermayer$^{\rm b,c}$, \vspace{1mm} \\
M.\ Pepe$^{\rm d}$, F.G.\ Rej\'on-Barrera$^{\rm a}$ and U.-J.\ Wiese$^{\rm b}$ 
\vspace{6mm} \\ 
\small $^{\rm a}$ Instituto de Ciencias Nucleares \vspace*{-2mm} \\ 
\small Universidad Nacional Aut\'{o}noma de M\'{e}xico \vspace*{-2mm} \\
\small A.\ P.\ 70-543, C.\ P.\ 04510 Distrito Federal, 
Mexico \vspace{0.2cm} \\
\small $^{\rm b}$ Albert Einstein Center for Fundamental 
Physics \vspace*{-2mm} \\
\small Institute for Theoretical Physics, Bern University \vspace*{-2mm} \\ 
\small Sidlerstrasse 5, CH-3012 Bern, Switzerland \vspace{0.2cm} \\
\small $^{\rm c}$ Institute for Theoretical Physics -- HAS, 
E\"{o}tv\"{o}s University \vspace*{-2mm} \\
\small P\'{a}zm\'{a}ny s\'{e}t\'{a}ny 1/a, 1117 Budapest, Hungary 
\vspace{0.2cm}\\
\small $^{\rm d}$ INFN, Sezione di Milano-Bicocca, Edificio U2 \vspace*{-2mm} \\
\small Piazza della Scienza 3, 20126 Milano, Italy\\ \ \\}

\begin{document} 

\date{}

\maketitle

\vspace{-1.2cm}

\begin{abstract} \normalsize

\noindent
We consider the 2d XY Model with topological lattice actions,
which are invariant against small deformations of the field 
configuration. These actions constrain the angle between neighbouring 
spins by an upper bound, or they explicitly suppress vortices (and 
anti-vortices). Although topological actions do not have a classical limit,
they still lead to the universal behaviour of 
the Berezinskii-Kosterlitz-Thouless (BKT) phase transition 
--- at least up to moderate vortex suppression.
Thus our study underscores the robustness of universality, which 
persists even when basic principles of classical physics are violated. 
In the massive phase, the analytically known
Step Scaling Function (SSF) is reproduced in numerical simulations. 
In the massless phase, the BKT value of the critical exponent 
$\eta_{c}$ is confirmed. Hence, even though for some topological actions 
vortices cost zero energy, they still drive the standard BKT transition. 
In addition we identify a vortex-free transition point, which
deviates from the BKT behaviour.

\end{abstract}

\newpage

\tableofcontents

\section{Introduction}

{\em Universality} is of central importance in quantum field theory and 
statistical mechanics, because it makes the long-distance physics 
insensitive to the short-distance details at the cut-off scale. The 
corresponding universality classes are determined by the space-time 
dimension and by the symmetries of the relevant order parameter fields. 
In lattice field theory, one often demands, in addition, the lattice 
action to have the correct classical continuum limit. 
Recently, we have introduced the concept of 
{\em topological lattice actions}, which do not have a classical 
limit \cite{Bie10}. Topological lattice actions are invariant 
against small deformations of the lattice fields. 
In $O(N)$ Models, the simplest topological action constrains the 
relative angle between nearest-neighbour spins to a maximal
angle $\delta$. All allowed configurations (that do not violate this 
constraint) are then assigned the action value zero. 
Since the action does not vary at all, 
it does not give rise to a meaningful classical 
equation of motion. Consequently, it does not have the correct classical 
continuum limit, and perturbation theory does not apply either. As we have 
demonstrated analytically for the 1d $O(2)$ and $O(3)$ Model, 
despite this classical deficiency, the topological lattice action
still leads to the correct {\em quantum continuum limit.} However, for 
these 1d topological actions the lattice artifacts go to zero only
as ${\cal O}(a)$ in the limit of vanishing lattice spacing 
$a$, while they are of ${\cal O}(a^2)$ for the standard lattice action. 

The correct quantum continuum limit has also been verified 
in the 2d $O(3)$ Model \cite{Bie10}. Based on numerical simulations 
with the Wolff cluster algorithm \cite{Wol89}, we have reproduced the 
analytic results for the Step Scaling Function (SSF) \cite{Bal04} 
that was introduced in Ref.~\cite{Lue91}. Remarkably, in the well
accessible range of correlation lengths, the cut-off effects of the 
topological action are {\em smaller} than those of the standard action 
and of the tree-level improved Symanzik action, which had been 
investigated previously \cite{Bal09}.
By combining the standard and the topological action, we have 
constructed a highly optimised constraint action for the 2d $O(3)$ 
Model that has only per mille level cut-off effects of the SSF
for ratios $a/L \leq 0.1$ \cite{Bal12}. Although the topological 
susceptibility receives contributions from zero-action dislocations,
it was found to diverge only logarithmically \cite{Bie10}, rather than with a 
power law, as a semi-classical argument would suggest \cite{Lue82}. While it 
has been suspected that $\theta$ is an irrelevant parameter which gets
renormalised non-perturbatively, we have identified distinct physical theories 
for each value $0 \leq \theta \leq \pi$ \ \cite{Boe11} 
(see also Refs.~\cite{Nogradi,dFPW}). At $\theta = 0$ we also 
investigated a topological lattice action which explicitly suppresses 
topological charges. Although this action does not have the correct 
classical continuum limit either, it was found to have the correct 
quantum continuum limit as well \cite{Bie10}.\\

This paper addresses the 2d XY (or $O(2)$) Model, 
which has been applied, for instance, to describe 
thin films of superfluid helium \cite{Minnhagen:1987zz}, 
fluctuating surfaces and their roughening transition, 
as well as Josephson junction arrays \cite{Sondhi:1997zz}. 
Here we investigate topological lattice actions for that model.
In contrast to the 2d $O(3)$ Model, which is asymptotically free, 
the continuum limit of the standard 2d XY lattice model is 
reached at finite values of the coupling. It corresponds to the 
well-known Berezinskii-Kosterlitz-Thouless (BKT) phase transition,
an {\em essential} transition of infinite order \cite{Ber70,Kos73}. 
The BKT transition separates a massive phase, in which vortices are
condensed, from a massless phase, with bound vortex--anti-vortex pairs.

Although it is not asymptotically free in the usual sense, 
the 2d XY Model has a non-trivial 
massive continuum limit at the BKT phase transition. There is 
numerical evidence that this continuum limit corresponds 
to the sine-Gordon Model at coupling $\beta \rightarrow \sqrt{8 \pi}$ 
\cite{Bal01}, which in turn is equivalent to the $SU(2)$ chiral 
Gross-Neveu Model. In this sense, the continuum theory is asymptotically 
free after all. The SSF \cite{Lue91} has been 
worked out analytically, and tested against 
numerical simulations \cite{Bal03}. Remarkably, in this case even the 
cut-off effects, which vanish only logarithmically as one approaches the 
continuum limit, have universal features \cite{Bal01a}.

It is interesting to investigate whether topological lattice actions 
lead to the usual quantum continuum limit also in this case.
One question is how far universality really reaches, in view of
the critical behaviour, and of the cut-off effects. 
As a further motivation, we refer to an estimate of 
the critical temperature for the standard lattice action, based
on the energy cost for isolated vortices (or anti-vortices),
which tend to disorder the system. If this is a relevant argument
behind the BKT phase transition, then the behaviour for topological
lattice actions is in fact tricky.

Some time ago, the BKT phase transition has been investigated in the 
so-called Step Model \cite{Gut73,Bar83,Nym86,San88}. 
The Step Model has a topological action, which vanishes if the angle 
between nearest-neighbour spins is less than $\pi/2\,$; otherwise it is 
a positive constant $S_0\,$.\footnote{Also the version
with a finite step at a variable angle has been addressed with analytical 
approaches \cite{Bar83,Nym86}.} While in the Step Model the BKT transition 
is attained by varying $S_0$, it is attained with the constraint action
by varying $\delta$. As $S_0$ is sent to infinity, the Step Model approaches 
the constraint action with $\delta = \pi/2$. On a square lattice, 
vortices are completely eliminated in that case. In agreement 
with the BKT picture, this point in the phase diagram turns out to 
be in the massless phase. For smaller values of $S_0$, vortices have a finite 
action. After some controversy, it has been confirmed that the Step Model is 
indeed in the BKT universality class \cite{Ken95,Ols01,Min03}. 

Using efficient cluster algorithms, we will show in this paper
that the constraint angle action also falls into the BKT universality class. 
This follows by comparison with analytic results for the SSF
\cite{Des92,Fev98}, and for the critical exponent $\eta_{c}$ 
\cite{Kos74,Jos77}. On the other hand, the cut-off effects of this 
topological action do not share the predicted universal features.

We further investigate a topological action that combines the constraint angle 
$\delta$ with explicit vortex suppression, by assigning an action value 
$\lambda > 0$ to each vortex or anti-vortex. Also that action turns 
out to have the universal features of the BKT transition, at least
up to $\lambda \approx 4$. A different behaviour is observed, however,
at the endpoint of this transition line,
which seems to be located at $\delta = \pi$ (no angle constraint) 
and $\lambda = \infty$ (no vortices).

In Section 2 we describe topological actions with two parameters,
for an angle constraint and an explicit vortex suppression.
Section 3 investigates these actions --- with the angle
constraint included --- by approaching the phase transition 
both in the massive and in the massless phase. In Section 4 we 
address a topological vortex suppression action without an angle 
constraint, and the extrapolation $\lambda \to + \infty$. 
Section 5 contains our conclusion. Finally the cluster algorithm 
used to simulate the topological actions is explained in Appendix A,
and Appendix B discusses surprising aspects of the correlations in
ferromagnetic systems.

\section{Topological Lattice Actions}

Let us consider the 2d XY Model on a periodic square lattice. 
A 2-component unit vector 
$\vec e_x = (\cos\varphi_x,\sin\varphi_x)$ is attached to each lattice 
site $x$. The standard lattice action reads
\begin{equation}
\label{standard}
S_{\text{standard}} [\vec e \, ] = \beta \sum_{\langle xy \rangle} \, 
[ 1 - \vec e_x \cdot \vec e_y ] = \beta 
\sum_{\langle xy \rangle} \, \Big[1 - \cos(\varphi_x - \varphi_y) \Big] \ ,
\end{equation}
where $\langle xy \rangle$ denotes a pair of nearest-neighbour sites, 
and the parameter $\beta$ corresponds to an inverse coupling. 
A vortex number $v_\Box \in \{0,\pm 1\}$ 
is associated with each elementary plaquette $\Box$, with the 
corners $x_1$, $x_2$, $x_3$, $x_4$ in counter-clockwise order. 
Introducing the relative angles
\begin{equation}
\Delta \varphi_{\langle x_i x_j \rangle} =  \left(\varphi_{x_i} - 
\varphi_{x_j}\right) \mbox{mod} \ 2 \pi \in \ (- \pi, \pi] \ , 
\end{equation}
the vortex number of a plaquette is given by
\begin{equation}
v_\Box = \frac{1}{2 \pi} \left(\Delta \varphi_{\langle x_1 x_2 \rangle} + 
\Delta \varphi_{\langle x_2 x_3 \rangle} + \Delta \varphi_{\langle x_3 x_4 \rangle} + 
\Delta \varphi_{\langle x_4 x_1 \rangle}\right) \in \{0,\pm 1\} \ .
\end{equation}
Higher vortex numbers cannot occur. The vortices are known to be the relevant 
degrees of freedom that drive the BKT phase transition \cite{Kos73}. According 
to Stokes' Theorem, the sum of all vortex numbers on a periodic lattice always 
vanishes, $\sum_\Box v_\Box = 0$. \\

Let us now introduce a topological action as a sum over elementary 
plaquettes,
\begin{equation}  \label{Slam}
S[\vec e\, ] = \lambda \sum_\Box |v_\Box| \ .
\end{equation}
This action counts the number of vortices (with $v_\Box = 1$) plus 
anti-vortices (with $v_\Box = - 1$), and multiplies this sum with 
the single-vortex action $\lambda$. In particular, the limit
$\lambda \rightarrow \infty$ removes all vortices.
When one continuously varies the spin field, without changing the 
(discrete) vortex number $|v_\Box|$, the action does not change either. 
Consequently, it is invariant against small deformations of the lattice 
field, so it represents a topological action. 

Let us mention that the analogous $\lambda$-term has also been introduced 
in the 3d XY Model \cite{KSW} and $O(3)$ Model \cite{LauDas}. 
In both cases it was combined with the standard term to investigate 
the phase diagram with the axes $\beta$ and $\lambda$. This also 
involved studies of the topological action at $\beta =0$, where
phase transitions at finite $\lambda_{c}$ were observed.

We may further modify the pure vortex suppression action by imposing 
the angle constraint $|\Delta \varphi_{\langle xy \rangle}| \leq \delta$, 
which restricts the relative angle $\Delta \varphi_{\langle xy \rangle}$ 
between nearest-neighbour spins $\vec e_x$ and 
$\vec e_y$ to a maximal value $\delta \in [0,\pi]$. 
Allowed configurations (which obey this angle constraint) still have the 
action value $S[\vec e\, ]$ of eq.\ (\ref{Slam}), 
while all other configurations (which violate 
the constraint on at least one nearest-neighbour pair of sites) are 
assigned an infinite action, so they are eliminated.
The actions characterised by the parameter $\lambda$ 
and the angle constraint $\delta$ remain invariant under 
small field deformations, and are thus still topological.

\section{Universal Behaviour of Angle Constraint Topological Actions}

In this section, we investigate the 2d XY Model with topological 
lattice actions that impose an angle constraint for nearest-neighbour 
spins, $\delta < \pi$. In addition, the actions may or may not 
explicitly suppress vortices, $\lambda \geq 0$. The universal 
behaviour is studied both in the massive and in the massless phase.

\subsection{Phase Diagram}

To determine the critical angle $\delta_c$ of the constraint topological 
action, we measure the correlation length $\xi(\delta)$ in the massive 
phase close to the phase transition that occurs in the infinite volume 
limit. This is done by increasing the lattice volume  $V = L\times L$ 
until the correlation length $\xi(\delta, L)$ converges to its infinite 
volume limit. For angles $\delta > \delta_c\,$, not too close to the 
phase transition, the convergence is observable on tractable lattice sizes 
(up to $L=2000$). To determine the critical point $\delta_c$ we fit the 
correlation length $\xi(\delta)$ to a function, which is characteristic 
for the BKT transition,
\begin{equation}
 \xi(\delta) = A \exp\left(B\left|
\frac{\delta_{c}}{\delta - \delta_{c}}\right|^{1/2}\right) \ ,
 \label{eq:crit}
\end{equation}
where $A$ and $B$ are fitting parameters. This form represents
an essential ({\it i.e.}\ infinite order) phase transition (for 
conventional lattice actions, the coupling $1/\sqrt{\beta}$ 
takes the r\^{o}le of $\delta$). The critical angles $\delta_c$ 
obtained from these fits (which have a good ratio $\chi^2/$d.o.f.) are 
listed in Table \ref{tab:dc} for the topological action without vortex 
suppression, $\lambda = 0$, and with explicit vortex suppression, 
$\lambda = 2$ and $\lambda = 4$.
\begin{table}[htb]
 \begin{center}
  \begin{tabular}{|r|r@{.}l|}
   \hline
   $\lambda$ & \multicolumn{2}{c|}{$\delta_c$} \\
   \hline
   \hline
   $0$ & 1&77521(57) \\
   $2$ & 1&86648(81) \\
   $4$ & 1&9361(83)  \\
   \hline
  \end{tabular}
 \end{center}
\caption{\it Critical angles $\delta_c$ for different topological 
actions, with vortex suppressing parameter $\lambda = 0, \, 2$ 
and $4$, based on fits to the function (\ref{eq:crit}).}
\label{tab:dc}
\end{table}

\begin{figure}[h!]
\begin{center}
\includegraphics[angle=0,width=.67\linewidth]{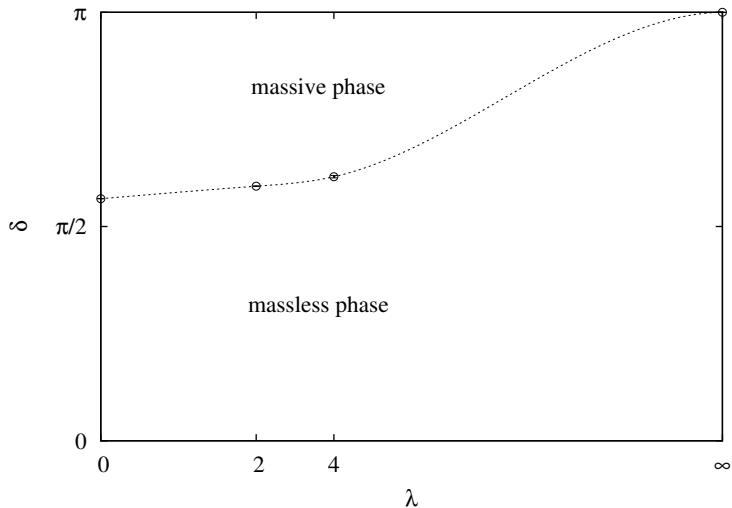}
\end{center} 
\vspace*{-5mm}
\caption{\it{A schematic illustration of the phase diagram, as expected 
based on the results for $\delta_{c}(\lambda)$ in Table \ref{tab:dc},
and anticipating the outcome of Section 4.}}
\label{phasedia}
\end{figure}

This suggests a phase diagram as sketched in Figure \ref{phasedia}. 
We expect the endpoint of the transition line to be located
at $(\lambda , \delta) = (+ \infty , \pi)$, see Section 4.

\subsection{Continuum Limit in the Massive Phase}
\label{sec:continuum_limit}

In order to investigate the continuum limit in the massive phase,
we consider the step-2 SSF \cite{Lue91}
\begin{equation}
\Sigma(2,u,a/L) =4L m(2 L) \ .
\end{equation}
Here $u =2 L m(L)$, and $m(L)$
is the size-dependent mass gap. Based on the exact S-matrix of the 
sine-Gordon Model, the SSF has been worked out analytically in the
continuum limit $\sigma(2, u) = \Sigma(2, u, a/L \rightarrow 0)$ \ 
\cite{Des92,Fev98}. Using the standard action, this analytic result has 
been confirmed in numerical simulations \cite{Bal03}. 
We mentioned before that here 
even the cut-off effects were predicted to have universal 
features. This refers to a lattice SSF of the form
\begin{equation}
\label{cutoff}
\Sigma(2, u, a/L) = \sigma(2, u) + \frac{c}{[\log(\xi/a) + U]^2} + 
{\cal O} \left( \frac{1}{\log^4(\xi/a)} \right) \ ,
\end{equation}
where $\xi = 1/m(L \rightarrow \infty)$ is the correlation length in
infinite volume. 

\begin{figure}[htb]
\begin{center}
\includegraphics[angle=0,width=0.9\linewidth]{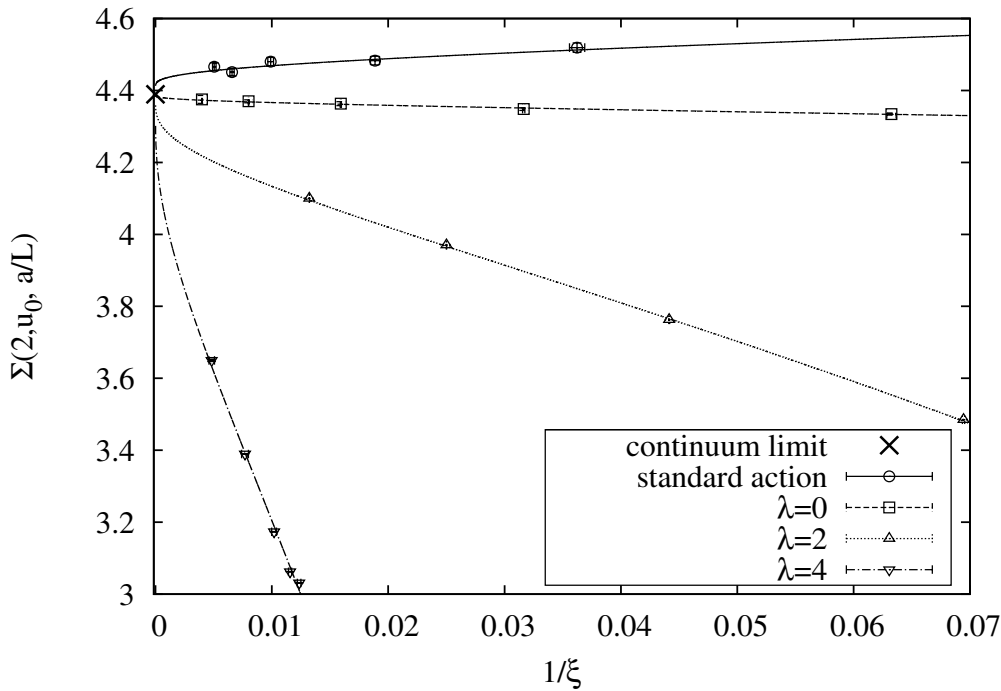}
\caption{\it Cut-off effects of the SSF $\Sigma(2,u,a/L)$
at $u = 3.0038$ for the standard action (data from 
Ref.~\cite{Bal03}), for the topological action without vortex 
suppression, $\lambda = 0$, and with explicit vortex suppression, 
for $\lambda = 2$ and $\lambda = 4$. All curves are fits to 
eq.~(\ref{cutoff}), where we insert the continuum limit
$\sigma(2,u) = 4.3895$.} 
\label{stepscaling}
\end{center}
\end{figure}

Figure \ref{stepscaling} illustrates the cut-off effects of the SSF
at $u = 3.0038$ for the standard action, and for the constraint 
topological action with the vortex suppression parameter
$\lambda = 0\,$, 2 or 4. The curves are fits to eq.~(\ref{cutoff}),
where we have inserted the analytically predicted
continuum SSF $\sigma(2,u)$ of Ref.~\cite{Bal03}.
As in the case of the 2d $O(3)$ Model, the standard action approaches 
the continuum limit from above, whereas the topological actions 
approach it from below. The continuum limit of the step-2 SSF
at $u = 3.0038$ amounts to $\sigma(2, u) = 4.3895$,
and the cut-off parameter $c = 2.618$ \cite{Bal03}
was supposed to be universal. 
In Table \ref{tab:ssf} we list our results, obtained by fitting
the parameters $\sigma(2,u),\ c$ and $U$ to the lattice data. 
The data for the standard action are taken from Ref.~\cite{Bal03}, 
where only $\sigma(2,u)$ and $c$ were fitted, since $U = 1.3$ 
is known from perturbation theory. These results indicate that all 
different actions converge to this continuum limit. For the 
topological actions, however, the fits yield negative and 
$\lambda$-dependent values for $c$. This suggests that 
both parameters, $U$ and $c$, depend on the lattice action and 
are therefore not universal.
\begin{table}[htb]
  \begin{center}
  \begin{tabular}{|c|r@{.}l|r@{}l|r@{.}l|r@{.}l|}
  \hline
   &\multicolumn{2}{c|}{$\sigma(2, u)$} & \multicolumn{2}{c|}{$c$} 
& \multicolumn{2}{c|}{$U$} & \multicolumn{2}{c|}{$\chi^2/$d.o.f} \\
   \hline
   \hline
 standard action & 4 & 40(2)   &    2.  & 4(6)   & 1    & 3      & 0 & 84 \\
   $\lambda = 0$ & 4 & 421(28) & $-4.$  & 0(3.6) & 4    & 1(2.2) & 0 & 15 \\
   $\lambda = 2$ & 4 & 427(23) & $-5.$  & 26(45) & $-0$ & 31(8)  & 2 & 51 \\
   $\lambda = 4$ & 4 & 71(25)  & $-21$ & (9) & $-0$ & 87(60) & 0 & 23 \\
   \hline
  \end{tabular}
 \end{center}
\caption{\it Fitting results for the cut-off effects of the SSF
in eq.~(\ref{cutoff}) for various lattice actions. 
The data for the standard action are taken from Ref.~\cite{Bal03}; 
they were obtained by fitting $\sigma(2, u)$ and $c$, 
whereas $U$ is known perturbatively. For the topological actions at 
$\lambda = 0, \ 2$ and $4$, we fitted $\sigma(2,u),\ c$ and $U$.}
\label{tab:ssf}
\end{table}

\begin{figure}[htb]
\begin{center}
\includegraphics[angle=0,width=0.9\linewidth]{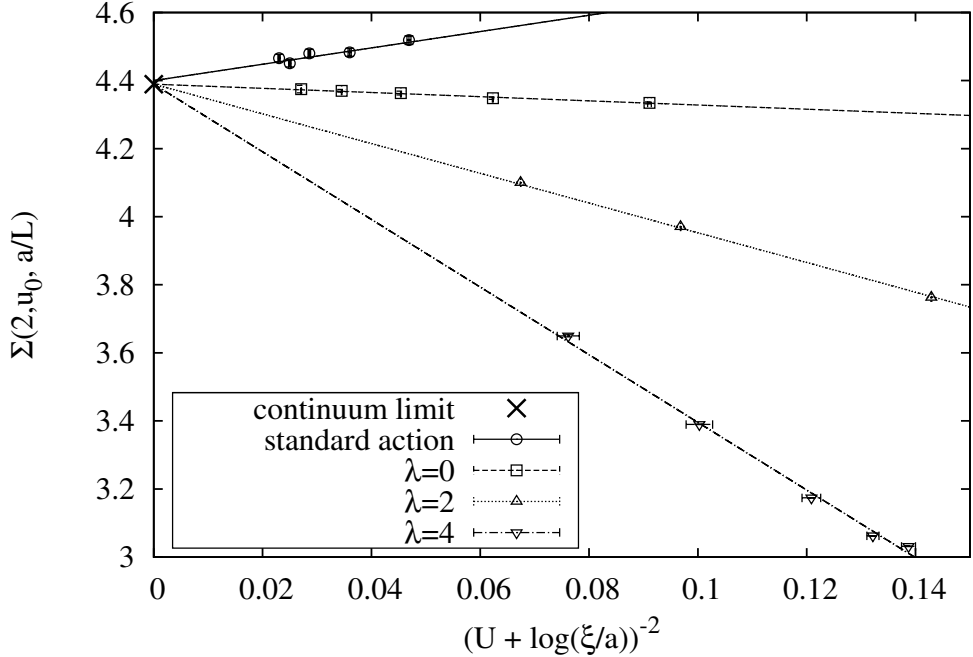}
\caption{\it Cut-off effects of the step-2 SSF 
$\Sigma(2,u,a/L)$ at $u = 3.0038$ for the standard action 
(data from Ref.~\cite{Bal03}), as well as for the 
topological action without vortex suppression, $\lambda = 0$, 
and with explicit vortex suppression, for $\lambda = 2$ and $\lambda = 4$. 
The curves are fits to  eq.~(\ref{cutoff}), where we have inserted 
the continuum limit $\sigma(2,u) = 4.3895$. The values on the
horizontal axis depend on the fitting parameter $U$, 
which is different for each action. Note that the plots in Figures
\ref{stepscaling} and \ref{stepscalinglog} contain (mostly invisible)
error bars in both directions.} 
\label{stepscalinglog}
\end{center}
\end{figure}

To illustrate the compatibility of our data with the analytic 
prediction, we follow Ref.~\cite{Bal03} and plot in 
Figure \ref{stepscalinglog} the same data as a function of 
$(U + \log(\xi/a))^{-2}$, where $\xi$ is still the infinite volume 
correlation length, and $U$ is a fitting parameter that differs for each 
action. In this plot we have again constrained the continuum limit 
of the SSF $\sigma(2, u)$ to its analytic prediction. 

\subsection{Critical Behaviour in the Massless Phase}

In contrast to second order phase transitions, only two critical 
exponents --- commonly denoted as $\eta$ and $\delta$ --- are 
defined in the conventional way also for the essential phase 
transition, which occurs in this model, cf.\ eq.~(\ref{eq:crit}).
Based on Renormalisation Group techniques, their values have been 
predicted to coincide with the corresponding exponents in the 2d 
Ising Model \cite{Kos74}. Here we focus on the exponent $\eta$, and 
its property to characterise the divergence of the magnetic 
susceptibility $\chi$.

The corresponding relation and the predicted critical 
value of $\eta$ are
\begin{equation}  \label{chieta}
\chi = \frac{1}{V} \left\langle 
\Big( \sum_{x} \vec e_{x} \Big)^{2} \right\rangle
\propto \left\{ \begin{array}{ccc}
\xi^{2 - \eta} && {\rm massive~phase} \\
L^{2 - \eta} && {\rm massless~phase}
\end{array} \right. \ , \qquad \eta_c = 1/4 \ ,
\end{equation}
in a square volume $V= L^{2}\,$. We now focus on the massless 
phase and insert the measured values of $\chi$ into the formula
\begin{equation}  \label{etaeq}
\eta = 2 - \frac{\ln (\chi / C)}{\ln L} \ ,
\end{equation}
where $C$ is the proportionality constant of eq.~(\ref{chieta}).
At least within the massless phase, {\it i.e.}\ for $\delta < \delta_{c}$,
it should be possible to find a constant $C$, which makes the results 
for $\eta$ in different volumes coincide to a good approximation \cite{FFS}. 

\begin{figure}[h!]
\vspace*{-5mm}
\begin{center}
\includegraphics[angle=0,width=.66\linewidth]{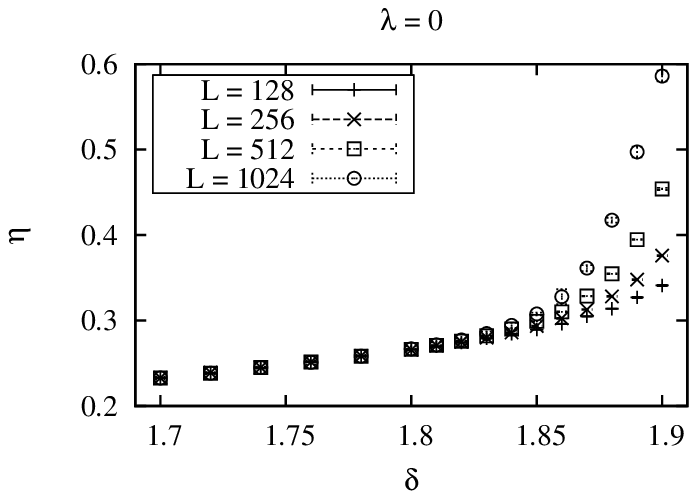}
\includegraphics[angle=0,width=.66\linewidth]{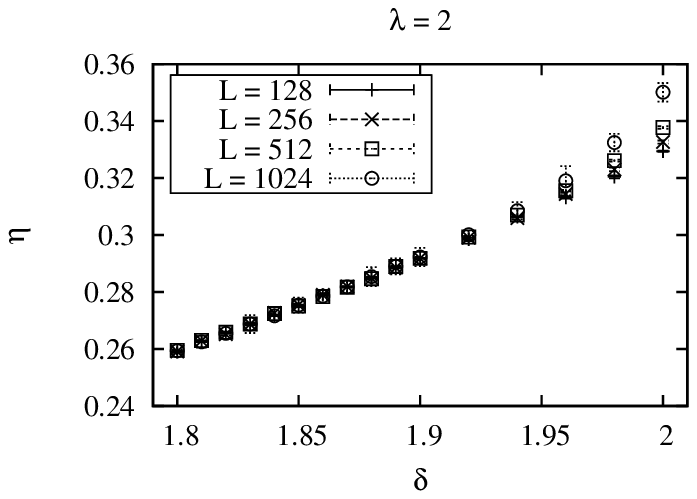}
\includegraphics[angle=0,width=.66\linewidth]{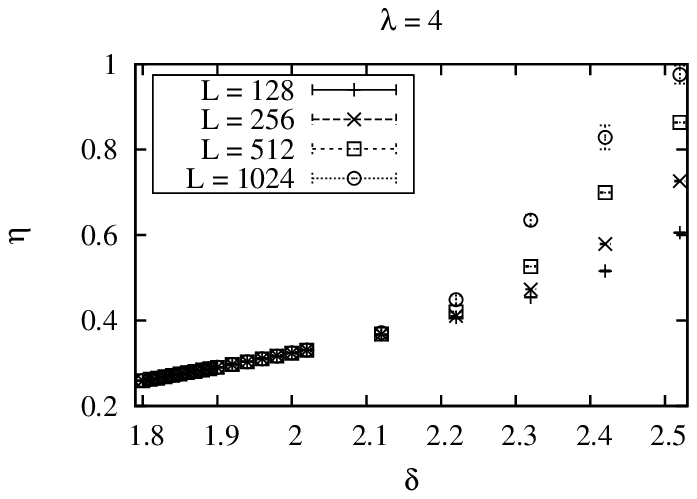}
\end{center} 
\vspace*{-5mm}
\caption{\it{The dependence of the exponent
$\eta$, according to eq.~(\ref{etaeq}), on the constraint 
angle $\delta$ at $\lambda = 0$, $2$ and $4$.}}
\label{etavsdelta}
\end{figure}

Figure \ref{etavsdelta} shows our results for $\lambda = 0$, $2$, 
$4$, and $L= 128, \dots ,1024$, with the optimal choice for the constant
$C$ at each $\lambda$. We see that the qualitative prediction of a 
coincidence of the $\eta$ values in different volumes, up to
some limiting angle $\delta_{\rm limit}\,$, is well confirmed. 
One is now tempted to interpret $\delta_{\rm limit}$ as an
estimate for $\delta_c$ \cite{FFS}. 
Table \ref{tabnaiv} shows that these values match
the expected magnitude, but they are significantly higher than the 
precise results for $\delta_c \,$, given in Table \ref{tab:dc}. 
Hence the coincidence of $\eta$ persists even in some (narrow) 
region of the massive phase (although eq.~(\ref{etaeq}) 
does not apply anymore).

\begin{table}[h!]
\centering
\begin{tabular}{|c||c|c|c|}
\hline
$\lambda$ & $0$ & $2$ & $4$ \\
\hline
\hline
$\delta_{\rm limit}$ & $1.825(5)$ & $1.93(1)$ & $2.17(5)$ \\
\hline
$\eta (\delta_c )$ based on eq.~(\ref{etaeq}) & 
$0.255(2)$ & $ 0.278(2)$ & $ 0.301(1)$ \\
\hline
\end{tabular}
\caption{\it Results for the limiting angle $\delta_{\rm limit}$
for the coincidence of the $\eta$ values in different volumes,
and for the critical exponent $\eta_c$ obtained from relation (\ref{etaeq}).}
\label{tabnaiv}
\end{table}

If we na\"{\i}vely extract the $\eta$-values at $\delta_{c}$, we obtain 
results for $\eta_c$, which are again in the predicted magnitude, but 
without a satisfactory precision, see Table \ref{tabnaiv}.
The $\eta_c$ values determined by this simple method tend to 
be too large, in particular for sizable $\lambda$ values.

Similar problems are notorious in numerical
studies of the standard action, the Villain action,
the Step Model etc.\footnote{A direct consideration of the
correlation function $\langle \vec e_x \vec e_{x+r} \rangle
\propto r^{-\eta}$ is plagued with even worse practical problems.}
The situation improves as one includes a {\em logarithmic
correction} to the finite size behaviour of $\chi$, which
has also been elaborated analytically in 
Ref.~\cite{Kos74},\footnote{On the other hand, this logarithmic 
correction term hardly affects the plots in Figure \ref{etavsdelta}.}
\begin{equation}  \label{chietar}
\chi \propto L^{2 - \eta} ( \ln L)^{-2r} \ , \qquad r_c = - 1/ 16 \ .
\end{equation}
Much of the literature that dealt with conventional lattice actions
focused on attempts to evaluate the critical exponent $r_c$
\cite{Ken95,PatSei,CPRV,Janke,HasPin,JasHah,DMC,ChaStr}.
Its numerical measurement is extremely difficult, 
as expected for a small exponent of a logarithmic term.
An overview of the results on this long-standing issue is given
in Ref.~\cite{Kenna}. Only in 2005 Hasenbusch reported a value 
which seems to confirm the prediction decently, $r_c = -0.056(7)$ 
\cite{MHas}. However, in his study of the standard action on 
lattices up to size $L=2048$, Hasenbusch had to fix $\eta_c = 1/4$ as
an input, and to introduce yet another free parameter by extending 
the logarithmic factor to $({\rm const.} + \ln L)^{-2r}$.

We first try to estimate the exponents $\eta_c$ and $r_c$
by fitting our data on lattice sizes $L = 128, \dots ,1024$
measured at $\delta$ angles slightly above and below $\delta_c \,$.
The fits have a good quality, and the results are given in 
Table \ref{tabetar}. The theoretical value $\eta_c = 1/4$ is 
reproduced well at $\lambda =0$ and approximately at $\lambda =2$.
However, at $\lambda =4$ we obtain an $\eta_c$ value which is clearly 
too large. Nevertheless this is compatible with the scenario that
the topological actions considered here are in the BKT universality 
class, and that the finite size effects are amplified for increasing 
$\lambda$ --- in qualitative agreement with the 
observations of Subsection \ref{sec:continuum_limit}.
Since a sizable $\lambda$ value suppresses the vortex density, it takes
a very large volume to provide a sufficient number of vortices to
drive an (approximate) BKT transition --- in line with the
picture of Ref.~\cite{Kos73}.

Regarding the logarithmic term in eq.~(\ref{chietar}), we do obtain 
small exponents of $| r_c | = {\cal O}(0.1)$ or below, but within 
this magnitude we cannot reproduce of the exact prediction.\\

\begin{table}[h!]
\centering
\begin{tabular}{|c|c||c|c|c|}
\hline
$\lambda$ & $\delta$ & $\eta$ & $r$ & $\chi^{2}$/d.o.f \\
\hline
\hline
\multirow{2}{*}{$0$}
& $1.76$ & $0.2563(66)$ & $-0.016(19)$ & $0.022$ \\
& $1.78$ & $0.2446(68)$ & $-0.034(19) $ & $0.05$ \\
\hline
\multirow{2}{*}{$2$}
& $1.86$ & $0.255(26)$ & $~~0.060(74)$ & $0.47$ \\
& $1.87$ & $0.2558(15)$ & $~~0.070(14) $ & $0.11$ \\
\hline
\multirow{2}{*}{$4$}
& $1.92$ & $0.366(11)$ & $-0.194(32)$ & $0.087$ \\
& $1.94$ & $0.317(25)$ & $-0.0470(66) $ & $0.012$ \\
\hline
\end{tabular}
\caption{\it Results for the determination of the exponents
$\eta$ and $r$ in eq.~(\ref{chietar}), by fitting our data 
at $L=128,\, 256, \, 512$ and $1024$, in the vicinity of 
the critical points.}
\label{tabetar}
\end{table}

\begin{figure}[h!]
\begin{center}
\includegraphics[angle=0,width=.59\linewidth]{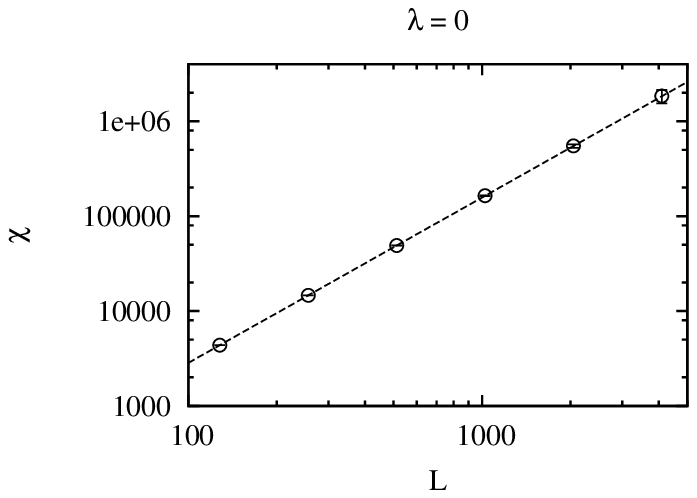}
\includegraphics[angle=0,width=.59\linewidth]{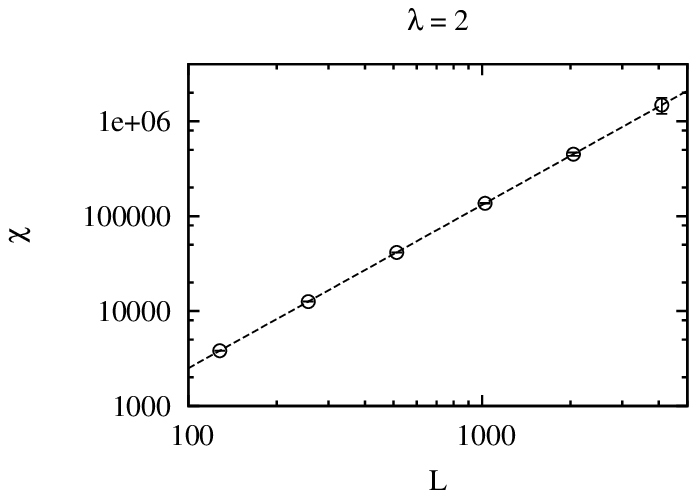}
\includegraphics[angle=0,width=.59\linewidth]{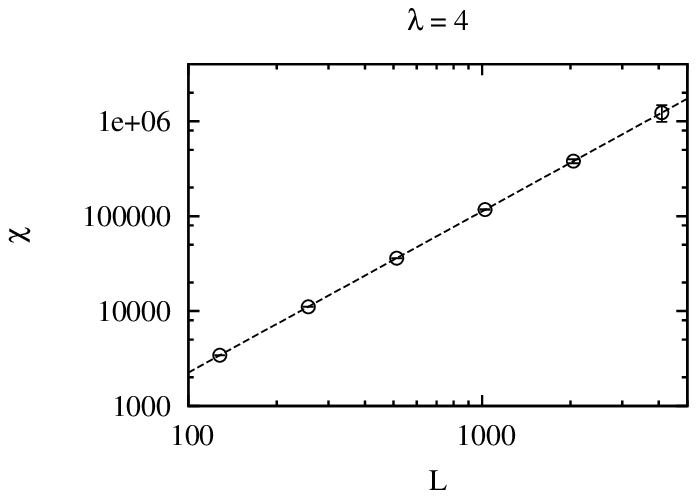}
\end{center} 
\vspace*{-5mm}
\caption{\it{Numerical results for the susceptibility $\chi$, measured 
at the critical points for $\lambda = 0$, $2$ and $4$, on lattices
of size $L=128, \dots , 4096$. The fits refer to eq.~(\ref{chietarr})
with fixed exponents $\eta_c = 1/4$, $r_c = - 1/16$, and $a_1$,
$a_2$ as free parameters. Here and in Table \ref{tabfreea}
we see that these fits are accurate in all three cases, confirming
the compatibility of our data with a BKT phase transition.}}
\label{plotfreea}
\end{figure}

Motivated by the strong finite size effects in this model,
Refs.~\cite{Ken95,Janke} worked out even a sub-leading 
logarithmic correction, which extends ansatz (\ref{chietar}) to
\begin{equation} \label{chietarr}
\chi = L^{2 - \eta} ( \ln L)^{-2r} \Big( a_1 + a_2 
\frac{\ln ( \ln L )}{\ln L} \Big) \ ,
\end{equation}
where $a_1$ and $a_2$ are constants. We add fitting results
to this extended formula, based on our data
measured at $\delta_c$ with fixed exponents
$\eta_c = 1/4$, $r_c = -1/16$, such that only $a_1, \, a_2$ are
free parameters. Table \ref{tabfreea} and Figure \ref{plotfreea}
show that these data match this form accurately for $\lambda =0$,
$2$ and $4$, if we consider some range with $L \geq 128$.\footnote{Table
\ref{tabfreea} also shows that the ratio $|a_2 / a_1|$
increases rapidly with $\lambda$. Hence the supposedly sub-leading
term in eq.~(\ref{chietarr}) dominates more and more,
which is consistent with the previous observation that
finite size effects are very strong at $\lambda =4 $.}
This observation provides satisfactory evidence that the behaviour
in these points is compatible with the BKT characteristics,
so that the topological actions do belong to the standard universality 
class, in agreement with Subsection \ref{sec:continuum_limit}. 
We assume this behaviour to persist for all points on the 
transition line with $0 \leq \lambda \lsim 4$. The limit
$\lambda \rightarrow + \infty$ will be addressed in the next section.

\begin{table}[h!]
\centering
\begin{tabular}{|c|c||c|c||c|}
\hline
& $L_{\rm min}$ & $a_1$ & $a_2$ & $\chi^2$/d.o.f. \\
\hline
\hline
\multirow{5}{*}{$\lambda =0, \ \delta = 1.77521$} 
& ~$32$ & ~~\,0.393(11)   & 1.056(33) &  9.422 \\
& ~$64$ & ~~\,0.4175(45)  & 0.981(13) &  0.801 \\
& $128$ & ~~\,0.4338(28)  & 0.9301(89)  &  0.084 \\
& $256$ & ~~\,0.4465(45)  & 0.888(15)  &  0.028 \\
& $512$ & ~~\,0.4753(69)  & 0.789(24) &  0.004 \\
\hline
\multirow{5}{*}{$\lambda =2, \ \delta = 1.86648$} 
& ~$32$ & ~~\,0.117(14) & 1.621(41) & 13.83 \\
& ~$64$ & ~~\,0.1550(83)  & 1.503(25) &  2.02 \\
& $128$ & ~~\,0.1829(16)  & 1.416(5)  &  0.022 \\
& $256$ & ~~\,0.1766(26)  & 1.437(8)  &  0.009 \\
& $512$ & ~~\,0.1618(46)  & 1.488(16) &  0.002 \\
\hline
\multirow{5}{*}{$\lambda =4, \ \delta = 1.9361$} 
& ~$32$ & $-0.073(18)$ & 1.999(52) & 25.45 \\
& ~$64$ & $-0.0232(68)$  & 1.847(20) &  1.548 \\
& $128$ & $-0.0005(36)$  & 1.774(12)  &  0.131 \\
& $256$ & ~~\,0.0128(53)  & 1.730(18)  &  0.050 \\
& $512$ & ~~\,0.0406(98)  & 1.634(33) &  0.014 \\
\hline
\end{tabular}
\caption{\it{Fitting results for the data at the critical
angle $\delta_c$, in the range  $L_{\rm min}$ to $L_{\rm max} = 4096$.
We fit the magnetic susceptibility $\chi$
to eq.~(\ref{chietarr}), with the predicted 
critical exponents $\eta_c = 1/4$, $r_{c}= - 1/16$.
For  $L_{\rm min} \geq 128$ the fits work very well,
which confirms the compatibility of our data with the critical
behaviour of the BKT universality class of the 2d XY Model.}} 
\label{tabfreea}
\end{table}

\section{Continuum Limit of the pure Vortex Suppression Action}

\begin{figure}[htb]
\begin{center}
\includegraphics[angle=0,width=0.8\linewidth]{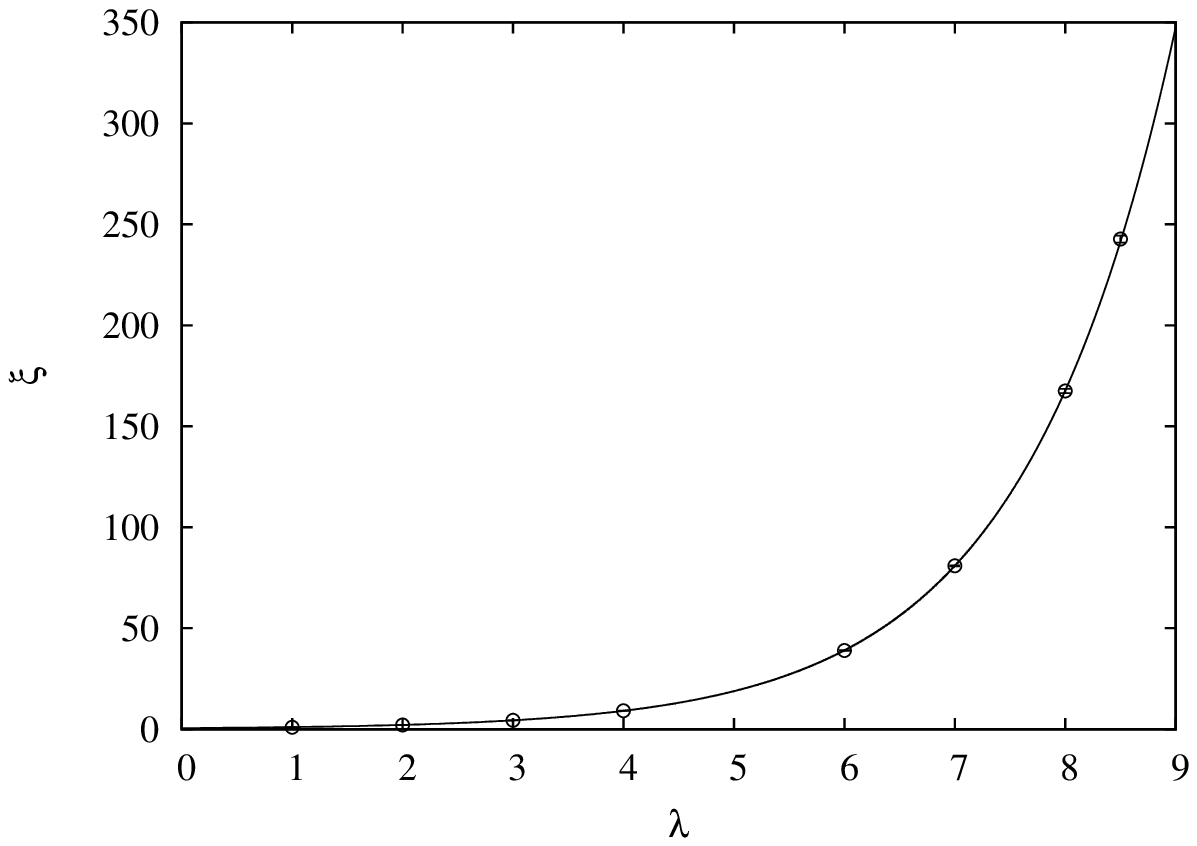}
\caption{\it Correlation length $\xi$ on large lattices as a function 
of the vortex suppression parameter $\lambda$. The fit to the exponential
function (\ref{expab}) 
(with $a = 0.492$, $b = 0.729$, $\chi^{2}/{\rm d.o.f.} = 0.290$)
indicates an essential phase transition at $\lambda = + \infty$.}
\label{fig:xi_lambda}
\end{center}
\end{figure}

We now investigate the vortex suppression action {\em without}
an angle constraint (which corresponds to $\delta = \pi$). Thus we 
consider the upper axis in the phase diagram of Figure \ref{phasedia}.
We have determined the infinite volume correlation length $\xi$ 
as a function of the vortex suppression parameter $\lambda$ on 
lattice sizes up to $V=2000\times 2000$. The results can 
be fitted well to the function 
\begin{equation}  \label{expab}
\xi(\lambda) = a \exp\left(b \lambda \right) \ ,
\end{equation}
where $a$ and $b$ are fitting parameters, see Figure \ref{fig:xi_lambda}. 
This suggests that the critical value is at $\lambda = + \infty$, 
as we anticipated in Figure \ref{phasedia}. This limit can be viewed
as a {\em plaquette constraint action.}

Studying the transition by measuring the step-2 SSF 
\begin{equation}
\sigma(2, u) = \lim_{a\rightarrow 0} \, \Sigma(2, u , a/L)
\end{equation}
(cf.\ Subsection \ref{sec:continuum_limit}) confronts us with an
additional limitation. 
The numerical results show that for this action the finite 
size effects constrain the finite volume correlation length 
to $\xi(L) \lesssim  0.4 \,  L$. 
This restricts the range of the variable $u = 2m(L) L = 2L / \xi(L)$ 
to a regime $u \gtrsim 5.0$.

A restriction of this kind is natural in models with discrete
energy eigenvalues $\propto 1/L$ in a UV conformal limit 
\cite{JBpriv}.\footnote{This is the ordinary case; asymptotically
free theories (in the usual sense) are the exception, where 
any $u \in \R_{+}$ is possible.}
Also for the standard action in the 2d XY Model there 
is an upper bound 
\begin{equation}
\frac{\xi(L)}{L} \leq \frac{4}{\pi} +  
{\cal O} \left( \frac{1}{\log L} \right)
\end{equation}
in the massive phase, see {\it e.g.}\ Ref.~\cite{MHas} and references 
therein. 
Qualitatively, such an upper bound can be understood using inequalities
for ferromagnetic systems. This is briefly discussed in Appendix B.

\begin{figure}[htb]
\begin{center}
\includegraphics[angle=0,width=0.9\linewidth]{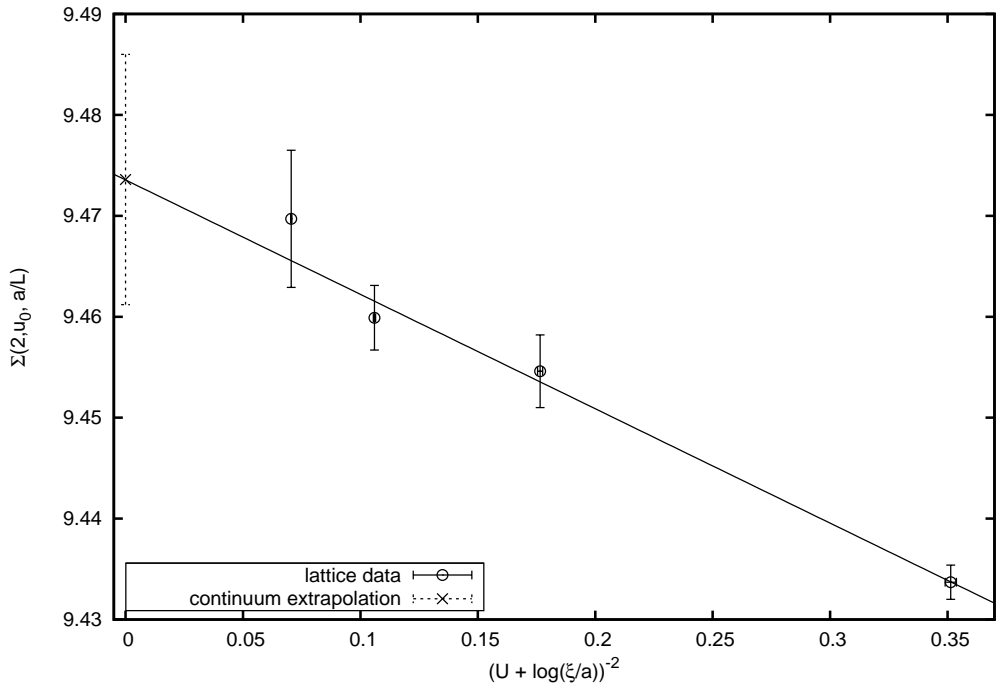}
\caption{\it Numerical data for the step-2 SSF $\Sigma(2, u, a/L)$ 
at $u = 6$, for the pure vortex suppression action,
fitted to eq.~(\ref{cutoff}). The parameters are given in
Table \ref{tab:ssf_lambda}. The continuum extrapolation
$\sigma(2, u)_{\rm fit}$ does not agree with the BKT value.}
\label{fig:ssf_lambda}
\end{center}
\end{figure}

We can still measure the step-2 SSF for $u$ sufficiently large,
for instance $u = 2m(L)L = 6$, and try to fit the cut-off 
behaviour with the function from eq.~(\ref{cutoff}), 
which describes the continuum limit at a BKT point. 
This fit, shown in Figure \ref{fig:ssf_lambda}, works quite
well. However, its continuum extrapolation
$\sigma(2, u )_{\rm fit} = 9.474(12)$, given in Table \ref{tab:ssf_lambda}, 
is rather far from the analytic BKT value of $\sigma(2, u) = 11.5314$ 
\cite{JBpriv} (which is close to $2 u=12$).
This suggests that the endpoint of the transition line
does {\em not} represent a BKT phase transition. Indeed, this
point is specific in the sense that one cannot cross it
(on the axis $\delta = \pi$).
Moreover, this observation is fully consistent 
with the established pictures of vortices driving the BKT transition 
\cite{Kos73}, so it cannot occur in the absence of vortices.
\begin{table}[htb]
 \begin{center}
 \begin{tabular}{|c||c|c|c||c|}
  \hline
$\chi^2/$d.o.f. & $c$ & $U$ & $\sigma(2, u)_{\rm fit}$ &
$\sigma(2, u)$ \\
   \hline
0.72 &  $-0.11(10)$ & 0.44(68) & 9.474(12) & 11.5314 \\
   \hline
   \end{tabular}
 \end{center}
\caption{\it Fitting result for the cut-off effects of the SSF 
$\Sigma(2, u, a/L)$ at $u = 6$, according to eq.~(\ref{cutoff}),
for the pure vortex suppression action.
The fitted continuum extrapolation $\sigma(2, u)_{\rm fit}$
does not agree with the BKT value $\sigma(2, u)$.}
\label{tab:ssf_lambda}
\end{table}

In the 3d XY Model, the analogous point 
($\lambda = + \infty$, with no other restriction) has been studied in
Ref.\ \cite{Fisch}. Also in that case the observation of the
transverse susceptibility $\propto L^{0.8}$ did not clarify
the properties of this vortex-free case.

\section{Concluding Discussion}

In this paper, we have investigated topological lattice actions for 
the 2d XY Model. At the classical level, these actions do not define 
a proper field theory, and perturbation theory is not applicable.

In order to efficiently simulate topological actions, we have employed 
variants of the Wolff cluster algorithm. Its application to the 
constraint action is straightforward, and for the vortex suppression 
action a generalisation to 4-spin interactions has been developed 
and applied successfully, see Appendix A. 

Despite its classical deficiencies, just as in the 2d $O(3)$ Model, 
we found that --- up to moderate vortex suppression --- 
topological actions yield the correct quantum continuum limit, 
which is here associated with the BKT phase transition. 
This includes in particular topological actions where 
vortices do not cost any energy. 
This observation is remarkable in light of attempts to derive the
critical line from the energy requirement for isolated vortices.

Specifically, in the {\em massive phase,} just as for the standard lattice 
action, the continuum limit is related to the sine-Gordon Model.
In the {\em massless phase} we have verified 
the usual BKT behaviour of the critical exponent $\eta_{c}$ --- and of
its logarithmic correction term --- with the topological actions. 
Our study demonstrates again the immense robustness 
of universality in quantum field theory, which does not rely on 
classical concepts.

An exception is the endpoint of this critical line, which seems to 
be located at $(\lambda , \delta ) = (+ \infty , \pi)$.
The extrapolation to this point --- which represents a plaquette
constraint action --- does {\em not} coincide with the BKT behaviour.
This agrees with the established picture that vortices (which are 
completely eliminated at this point) are required to arrange for a 
BKT transition \cite{Kos73}.\\

For comparison, we mention the case of the so-called 
Extended XY Model, with the lattice action \cite{DSS}
\begin{equation}
S [ \varphi ] = \beta \sum_{\langle xy \rangle} \Big[ 1 -
\cos^{2q}((\varphi_{x} - \varphi_{y})/2) \Big] \ ,
\end{equation}
in the notation of eq.~(\ref{standard}), and with $q > 0$. 
For $q=1$ it is equivalent to the standard action, 
but increasing $q$ leads to a more and more narrow
potential well for $\varphi_{x}-\varphi_{y}$, with
width $\approx \pi / \sqrt{q}$. The motivation was also
an explicit vortex suppression; in the quadratic approximation
to the potential they cost energy $\approx \beta q /2$.

Due to the gradual suppression of the vortices for 
increasing exponents (even without fully excluding them), 
Ref.\ \cite{DSS} predicted the phase transition to 
turn into first order above some value of $q$, so it would 
match the behaviour which is observed experimentally for melting 
films of noble gases adsorbed on graphite.

This Extended XY Model has been investigated in numerous
papers. The essential BKT phase transition is observed at
low values of $q$, and for some time the conjectured first order
transition at large $q$ was controversial. However, it is
now well confirmed numerically at $q \gsim 8$ \cite{Ota,SinhaRoy1}.
Moreover, an analytical proof for this conjecture 
was given in Ref.\ \cite{EntShlos}.
Ref.\ \cite{SinhaRoy2} added a vortex eliminating
term with $\lambda \to + \infty$ also in this case.
No phase transition was observed at finite $\beta$,
hence the authors concluded that not only the BKT transition,
but also the first order transition at large $q$ is driven by vortices. 

In contrast, for the Step Model no non-BKT phase transition has ever
been found, and for the topological lattice actions we do not observe any 
finite order transition in the $\delta$-$\lambda$ phase diagram either.
However, a change to first order along the transition line --- at
some large value of $\lambda$ --- is conceivable in our case as well
(that would not contradict universality).
If this occurs as in the Extended XY Model, then it should 
change again at the endpoint, according to Ref.\ \cite{SinhaRoy2}.

In any case, the characteristics of the transition at the vortex-free 
endpoint is an open question, to be explored in the future.\\

\vspace*{5mm}

\noindent
{\bf Acknowledgements :} 
We thank for
very useful communications with J.\ Balog, E.\ Seiler, P.\ Weisz and 
U.\ Wolff. This work was supported in parts by the {\it Schweizerischer 
Nationalfonds} (SNF), and by the {\it Consejo Nacional de Ciencia y 
Tecnolog\'{\i}a} (CONACyT), project 155905/10 ``F\'{\i}sica de 
Part\'{\i}culas por medio de Simulaciones Num\'{e}ricas''. 
The ``Albert Einstein Center for Fundamental Physics'' at 
Bern University is supported by the ``Innovations- und Kooperationsprojekt 
C-13'' of the Schweizerische Uni\-ver\-si\-t\"ats\-kon\-fe\-renz (SUK/CRUS).

\begin{appendix}

\section{Cluster Algorithm for the Vortex Suppression Action}

The algorithm for the vortex suppression angle constraint action is
based on the Wolff cluster algorithm \cite{Wol89}. In the single cluster
variant, each cluster update begins with the selection of an initial spin 
as a seed for cluster growth, and with the choice of a reflection 
line (a reflection hyper-plane in general $O(N)$ Models), which is 
perpendicular to the randomly selected unit vector $\vec r$. 
Starting with this seed, some spins $\vec e_x$ are combined to a cluster, 
which are then collectively reflected --- or {\em flipped} ---
to the new spin orientations 
$\vec e_x \, ' = \vec e_x - 2 (\vec r \cdot \vec e_x) \, \vec r$.
Spins may be put in the same cluster due to the nearest-neighbour angle
constraint, or due to the vortex suppression plaquette interaction. Two 
nearest-neighbour spins $\vec e_x$ and $\vec e_y$ are always put in the same 
cluster if the flip of $\vec e_x$ to $\vec e_x \,'$ (without flipping 
$\vec e_y$) would lead to a relative angle between $\vec e_x \, '$ and 
$\vec e_y$ beyond the constraint angle $\delta$. 

The cluster rules implied by the vortex suppression four-spin plaquette 
action are more complicated. Let us consider the spins $\vec e_{x_i}$ 
at the four corners $x_1$, $x_2$, $x_3$ and $x_4$ of a plaquette $\Box$, 
as well as their reflection partners $\vec e_{x_i} \, '$. 
Depending on whether a spin is flipped or not, there are 16 possible 
spin configurations on the given plaquette. Each one has a Boltzmann 
weight $\exp(- \lambda |v_\Box|)$, depending on the vortex number $|v_\Box|$ 
of the corresponding spin configuration. Since the reflection of all
four spins on a plaquette $\Box$ just changes the sign of the vortex charge 
$v_\Box$, each of the 16 spin configurations has a total reflection partner 
with the same Boltzmann weight. We can thus limit the discussion to 8 pairs 
of configurations. We distinguish two qualitatively different cases:
\begin{enumerate}
\item
In this simple case the vortex number is always zero,
irrespective of whether any spin is flipped or not.
Hence all 16 spin configurations have the same Boltzmann weight 1. 
Based on the vortex suppression action, there is no 
need to put any of these four spins in a common cluster.
\item
The second case can be characterised as follows: when all spins are 
flipped to the same side of the reflection line, the vortex number 
is necessarily zero. We denote this spin configuration as the
``reference configuration''. When each of the spins is individually
flipped (without flipping any other spins), there are two spins 
whose flip generates a vortex (or an anti-vortex). 
We denote these two as the ``active spins''. 
It turns out that the simultaneous flip of two spins (starting out of 
the reference configuration) generates a vortex only if exactly 
one of the two spins is active. If both or none of the two flipped 
spins are active, no vortex is generated. If three or four spins are 
flipped simultaneously, one just generates the total 
reflection partners of the previously discussed cases. 

This gives rise to the following cluster formation rule. If the two 
active spins are on the same side of the reflection line, they are 
put into the same cluster with probability $1 - \exp(- \lambda)$, 
otherwise they remain independent. The other spins are not affected 
by the vortex suppression action on this plaquette and remain 
independent. Still, preliminarily independent spins may finally become 
members of the cluster due to the angle constraint, or due to the vortex 
suppression action on a neighbouring plaquette.\footnote{It should be 
noted that two active spins that are tied together in the same cluster 
may actually end up not to belong to the single cluster that is currently 
being built. In any case, one must keep track of the plaquettes on which 
a decision based on $|v_\Box|$ has already been taken, and one must stick 
to that decision when this plaquette is visited again, in the process of 
identifying the cluster members.\label{fnalgo}} 
For the efficiency of the algorithm it is essential that spins are 
put in the same cluster only if they are on the same side of the 
reflection line. This prevents the clusters from becoming unphysically 
large (their linear size should be of ${\cal O}(\xi )$).

This algorithm obeys detailed balance. In particular, when a 
plaquette carries a vortex (and thus has the Boltzmann weight 
$\exp(- \lambda |v_\Box|) = \exp(- \lambda)$), 
the two active spins are {\em not} put in the same cluster
(with probability $w = 1$), because they are then necessarily on two 
different sides of the reflection line. On the other hand, if the 
two active spins are on the {\em same} side of the reflection line, 
$v_\Box' = 0$ and the Boltzmann weight is $\exp(- \lambda |v_\Box'|) = 1$. 
In that case, the two active spins are put in the same cluster with 
probability $1 - w' = 1 - \exp(- \lambda)$, while they remain 
independent with probability $w' = \exp(- \lambda)$. Only in the latter 
case, the two active spins may not belong to the same cluster, and are thus 
flipped independently, which again results in the creation of a vortex. 
Hence the detailed balance relation connecting the two configurations reads
\begin{equation}
\exp(- \lambda |v_\Box|) \, w = \exp(- \lambda) 
= \exp(- \lambda |v_\Box'|) \, w' \ .
\end{equation}
\end{enumerate}

As we have explicitly verified in an extensive computer search, other 
cases do not exist. Once spins have been put together in the same cluster 
(due to the nearest-neighbour angle constraint action, and/or due to the 
vortex suppression plaquette interaction), all spins $\vec e_x$ in the 
cluster are simultaneously flipped to $\vec e_x \, '$. Then a new random 
site is selected as a seed for cluster growth, along with a new unit 
vector $\vec r$, and the entire procedure is repeated. \\

As an alternative to this single-cluster algorithm, we also employed
a multi-cluster algorithm, which constructs all clusters in a spin 
configuration and flips each of them with a probability of $1/2$. 
Then the subtleties explained in footnote \ref{fnalgo} do not occur.

An additional virtue of cluster algorithms is the applicability of 
improved estimators. For the variant that updates the vortex 
suppression angle constraint action, the improved estimators --- 
for example for the correlation function and the susceptibility ---
work exactly as in the original Wolff algorithm \cite{Wol89}.

\section{On Inequalities for Ferromagnetic Systems}

Consider the standard action (\ref{standard}) on a long strip with
$N=L/a$ sites on a time-slice. (We take below $a=1$ for simplicity.) 
Making the ferromagnetic coupling anisotropic, 
$\beta \to (\beta_{x}, \beta_{t})$, we increase $\beta_{x} \to \infty$
while keeping $\beta_{t} = \beta$ constant. This way the 2d system 
turns into a 1d chain with $\beta' = N \beta$.
By increasing a $\beta$-parameter in a ferromagnetic system 
one might expect that the correlation length could only grow. 
This implies a lower bound for the correlation
length in the original model (with isotropic coupling)
\begin{equation}
\xi (\beta ; N) \le \xi_{1}(N \beta) = 2 \beta N + \mathcal{O}(1) \ ,
\end{equation}
where $\xi_{1} (\beta')$ is the correlation length for the 1d chain.

According to Ginibre's Theorem \cite{Ginibre} this intuitive 
argument indeed holds for the standard action, and for a large class 
of further actions specified in Ref.\ \cite{Ginibre}. \\

Surprisingly, for slightly more complicated actions this inequality
does not hold. Consider the nearest neighbour action with the action density
\begin{equation} \label{bgd}
  s(\evec, \evec{\, '}) = \beta (1 - \evec \cdot \evec{\, '}) 
  + \gamma (1 - \evec \cdot \evec{\, '})^2 
  + s_{\mathrm{constr}}(\evec \cdot \evec{\, '} - \cos \delta ) \ ,
\end{equation}
where the last term describes the constraint 
$\evec \cdot \evec{\, '} > \cos \delta$.
One can make the system ``more ferromagnetic'' by increasing 
$\beta_x$ or $\gamma_x$, or by decreasing $\delta_x$.
Taking again the 1d limit (say, by $\beta_x \to \infty$)
one would na\"{\i}vely expect 
\begin{equation}
\xi(\beta, \gamma, \delta ; N) \le \xi_{1} (N \beta , N \gamma , \delta) \ .
\end{equation}
This, however, cannot be true, since one has
$\xi_{1}(0, N \gamma, \pi ) \propto \sqrt{N \gamma}$ for 
$N \gamma \to \infty$ and $\xi_{1}(0,0,\delta) \propto 1/\delta^{2}$ 
for $\delta\to 0$, while the left-hand-side
increases $\propto N$ in the massless phase.
(Of course, the action \eqref{bgd} does not satisfy the conditions
of Ginibre's Theorem.)
\begin{table}[h!]
  \centering
  \begin{tabular}{|c|c|c|c||c|}
    \hline
    $\beta_x$ & $\beta_t$ & $\gamma_x$ & $\gamma_t$ & $\xi$  \\
    \hline
    \hline
    $0.0$ & $0.0$ & $1.0$ & $1.0$ &  $8.8316$ \\
    $0.0$ & $0.0$ & $1.1$ & $1.0$ &  $8.7980$ \\
    \hline
    $1.0$ & $1.0$ & $1.0$ & $1.0$ & $13.9148$ \\
    $1.0$ & $1.0$ & $1.1$ & $1.0$ & $13.9008$ \\
    $1.1$ & $1.0$ & $1.0$ & $1.0$ & $13.8877$ \\
    \hline
    $1.0$ & $1.0$ & $0.0$ & $0.0$ &  $6.4263$ \\
    $1.1$ & $1.0$ & $0.0$ & $0.0$ &  $6.4553$ \\
    \hline
  \end{tabular}
  \caption{\it The change of the correlation length 
    by increasing the spatial parameters 
    $\beta_x$ or $\gamma_x$ on a strip with $N=2$ sites.
    The last pair of data refers to the standard action
    where Ginibre's Theorem applies, so that the intuitive 
    expectation holds.}
  \label{bdtable}
\end{table}

In Table~\ref{bdtable} we illustrate this behaviour for the mixed
action (without the constraint, $\delta=\pi$), where the inequality 
is violated, and for the standard action ($\gamma=0$, $\delta=\pi$)
where it holds.

\end{appendix}

\end{document}